# Functional Categories of Support to Operations in Military Information Systems

(August 2000)

**Dr. Andreas Tolk**

Industrieanlagenbetriebsgesellschaft mbH
Einsteinstr. 20
85521 Ottobrunn
Germany

**Keywords**: Federated Solutions, C3I, OOTW, Reusable Components, Decision Support, Simulation, NATO COBP

**Summary**

In order to group the functional requirements for support to operations by modern information systems systematically, the NATO Code of best Practise (COBP) for C2 Assessment defines three domain areas: Battlespace Visualization, Decision Making, and Battle Management Functions. In addition, within an domain overlapping information grid of the information system, necessary functions for assessing and disseminating the information are capsulated. For all three domains, including the overlapping information grid, the respective requirements for functional support have to be met be future command, control, communications, and intelligence (C3I) systems.

The paper describes the functional categories of the three domains having been defined for article 5 operations, extends them to meet the requirements for operations other than war (OOTW), and gives some examples how modules of simulation systems can deliver respective support functions. In addition, references defining migration procedures for legacy systems to enable a smooth change from the old to the new C3I paradigm are given.

**Introduction**

The US Department of Defense (DoD) *Defense Information Infrastructure (DII) Master Plan, Version 8* [DISA, 1999] identifies the future military operations environment as including:
- Regional orientation
- Nontraditional, transitional, and unpredictable threats
- Ad-hoc coalitions and/or unilateral operations
- Adaptive planning and strategic agility
- Smaller total force – reduced forward presence
- Rapid response capability
- Military Operations Other than War (MOOTW, e.g., peacekeeping, sanctuary, etc.)
- Tailored force packages deployed under JTF (Joint Task Force) or CTF (Combined Task Force) command
- Reduced funding
- Asymmetric risks: terrorism, proliferation of weapons of mass destruction, and information warfare.

The United States (US) Department of Defense (DoD) vision for the future, as described in *Joint Vision 2010* [DoD, 1996], is one of information superiority. This vision was formulated with the recognition that operational decision making, in both combat and peacetime, will demand increasing rapidity and accuracy. This is due, in part, to the reduced force structure and budgets of the US military. It is also due to the fact that the military is being asked to expand its mission areas to include establishing and maintaining regional peace and stability, humanitarian relief, and disaster recovery. Additionally, the military finds itself in joint, combined, and coalition operations in support of these varied mission areas. This extended and still new tasks are much more demanding support by modern information system technology then ever before.

This paper wants to give some hints, how this challenge can be met and how new findings in recent R&D studies in the C4I section as well as in the simulation section can help in finding appropriate solutions. In order to do so, the paper is divided into two main sections.
- In the first one, the functionality of future C4I systems is introduced. The functionality domains and function groups are based on the findings of recent and ongoing NATO activities within the studies, analysis, and simulation (SAS) panel of the NATO Research and Technology Board (RTB).
- The second part is dealing with the future role of simulation systems and simulation functionality within this context. Computer Assisted Exercises (CAX) as well as Support to Operations as Decision Support Systems are the topic of this part of the paper.

Additionally, the paper will deal with the role of standardization in this context. The kernel idea is to use the concept of federated solutions based on reference architectures and common shared data models instead of using the traditional way of standardizing the interfaces in a hard way.

The ideas of this paper are well known in several international institutions, as the ideas already are comprised in the papers awarded by the Simulation Interoperability Standardization Organization (SISO) as well is in some NATO publications and CCRP papers.

**Functionality of Future C4I Systems**

The NATO Code of Best Practice for C2 Assessment has been written to support the analysts of command and control issues. It has been written by a Research Study Group (RSG) on Modeling of Command and Control (C2). Participating nations were Denmark, France, The United Kingdom, Canada, The Netherlands, Norway, Spain, Turkey, and The United States. In addition, the efforts were supported by the NATO Command, Control, and Consultation Agency (NC3A).

Originally planned to focus on procurement, force planing, and operational and tactical evaluation, the focus broadened during the study. The NATO CoBP now comprises chapters dealing with
- Human Factors and Organizational Issues,
- Development of respective Scenarios,
- Definition and Use of different Measures of Merit,
- Use of Tools (Models) and their Application,
- Recommendations and Conclusions

After having finalized the part of the study dealing with „Article V" operations, a second part started dealing with operations other than war. Not only the topic of the NATO CoBP broadened, but also the number of participating and contributing nations.

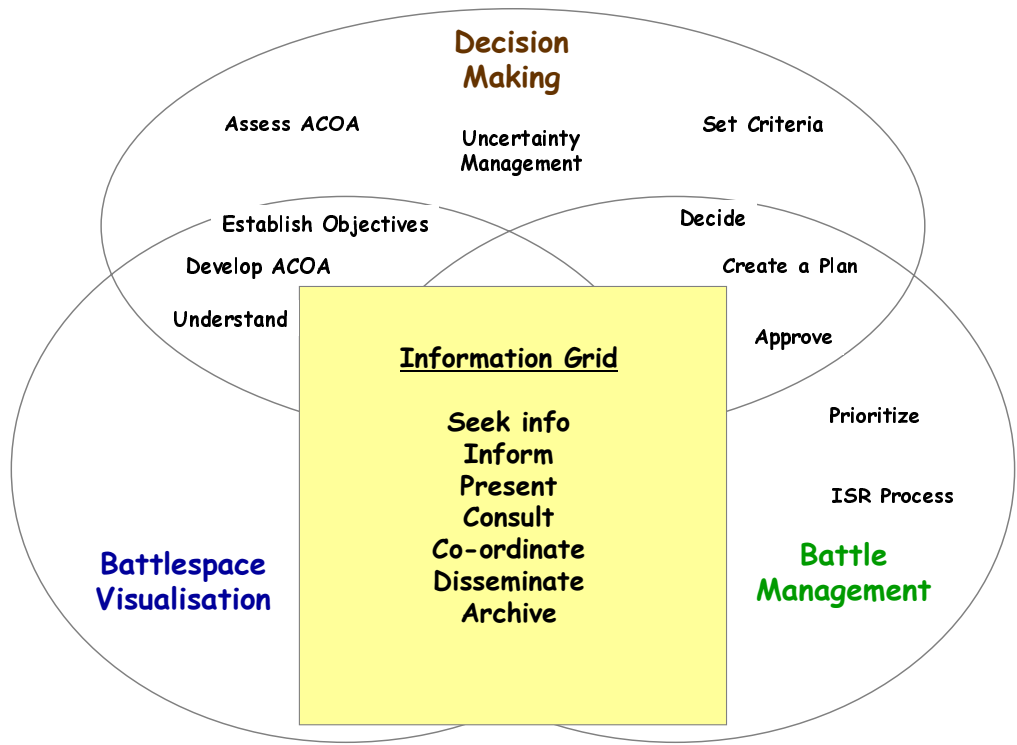

*Fig. 1*: Function Domains in „Article V" Operations

*Functionality for „Article V" Scenarios*

In this paper, the focus lies on the functionality to be supported by the respective C4I systems. Hence, the first subsection is going to introduce the functions to be supported for „Article V" operations, i.e., traditional defense operations.

Within the analyses having been conducted by the NATO experts during the CoBP works, three main domains of C4I functions have been identified:
- battlespace visualization,
- support of decision making, and
- battle management.

Figure 1 gives an overview of the three functionality domains, the connecting information grid as well as the comprised functions and function groups.

As can be seen easily, these functional domains correspond very well with the Command and Control process:

- **Understand**: The military decision maker must understand, what is going on on the battlefield. Objective is, to gain command and control superiority on the basis of information superiority.

- **Develop Alternative Courses of Action**: Based on the situation awareness gained, alternative courses of actions have to be developed. They comprise different orders of (potential) battle, equipment to be used, etc.

- **Establish Objectives**: What is also important is to define the own objectives to be reached in respective phases. These objectives have to be harmonized with the objectives of the superior command (What do I have to do?) als well as with the capabilities of the subordinated units (What can be done?)

- **Set Criteria**: In order to prepare the decision process, setting of decision criteria has to be supported. That doesn't mean that a utility based approach has to be used, or that any other form of rational decision making is enforced, but it just has to be assured that the relevant criteria are known by the decision maker. (If he is going to use this knowledge depends on him.)

- **Assess Alternative Courses of Action**: The ACOA having been developed earlier have to assessed in the planing and deciding process. In order to be able to do so, the results of the evaluation processes supported by the function groups "establish objectives" and "set criteria" have to be taken into account, preferably automatically.

- **Uncertainty Management**: As the perception of the military decision maker is unsharp, uncertain, incomplete and contradictive, respective uncertainty has to be taken into account, e.g., by using fuzzy sets or introducing probability intervals using the algorithms of the Dempster-Shafer-Theory etc. It should be pointed out that the objective is to manage the uncertainty to increase the situation awareness, not simply to reduce uncertainty.

- **Decide**: The decision is and remains the domain of the military decision maker. However, concerning that there are different levels of decisions [Tolk and Schiefenbusch, 1999; Tolk and Kunde, 2000], functions taking over routine work are needed to reduce the overall workload where possible.

- **Create a Plan**: The creation of the plan can be facilitated not only by integrated office automation tools, but by assessing the respective ACOA also. Many things being needed in the plan are already part of the ACOA, thus, why to do the same work twice?

- **Approve**: After the decision and creation of the plan, the process has to be harmonized staff-wide. In many cases, harmonization with the neighbors, the subordinates, etc., are also needed. This happens while approving the plan and has to be supported.

- **Prioritize**: Within the battle management domain, now the actions, resources, etc., have to be prioritized in the respective matter.

- **Information, Surveillance, and Reconnaissance Process**: Last but not least, the ISR process has to be harmonized also, opening with its result the next circle of command and control by delivering information, that has to be understood first, etc.

All three domains are interconnected by the respective C4I data domain, the "Information Grid" taking care of the (semi-) automated process of seeking the information within the potentially distributed data bases (including open sources like the Web), inform the military decision maker by presenting the results in the desired form to him (graphics, charts, slides, etc.), give him advice in form of consultations, coordinate the different cells and efforts, disseminate the results, information, reports and orders, and – last but not least – archiving the different steps for later analyses, consultation, or simply a debriefing.[1]

The functionality needed to support „Article V" operations by respective information technique is needed for operations other than war also. However, due to the special needs and demands, additionally new sets of functions have to be introduced. This is the topic of the next section.

*Functionality for Operations Other Than War*

Although already in „Article V" operations the need for a broader flexibility has been seen, it becomes a prior necessity in operations other than war (OOTW). A first analyses of new functions can be found in [Tolk & Kunde, 2000]. The inputs originates from the German work for the new NATO Code of Best Practice for Operations other than War.

First thing to mention is, that the function domains remain the same. There is no need to introduce new domains of functionality in OOTW that are not found in traditional scenarios as well. However, the function groups have to be much broader in many cases. These new functions, however, can again be useful in traditional scenarios also.

The first thing to cope with is the much broader information base. Information isn't restricted any longer to military information systems and sensors. ISR have to take into account other systems and open sources, including the Web, the Media, etc. It is impossible to ignore the influence of the media in OOTW, thus, it is crucial to have access to their sources, and – if this is not possible – at least to the media itself. The CNN monitor within the headquarter became a standard device in the recent years, and this effect will increase. Although the author doesn't know how to cope with this in simulations, at least in the training and exercise domain respective procedures have to find their way into the field manuals, and therefore into supporting training simulation systems. The challenge increases in the international environment, when several nations and doctrines have to be brought together.

---

[1] *The author is not claiming completeness. There are sure other functions that are important and any discussion is welcome at this point.*

Shoulder to shoulder with the traditional ACOA the concept of Alternative Contingency Plans (ACP) has to be seen within OOTW. Contingency plans are much more narrow in principle for the local military decision maker. However, looking at the latest military operations research findings, e.g. the so called "dominant maneuvers", the principle of preplanned operations or operation components finds its way into traditional military operation plans as well. New scenarios and tasks demands rapid, correct, and high qualified decisions from the military decision maker. Thus, support by information technique is a must, and the highest benefit from the "digitized battlefield" can be gained when IT systems help the warfighter to choose from a set of preplanned operation components.

Two new functions found their way into the decision domain also: Knowledge and Workflow Management. These both functions take the nature of OOTW into account by
- managing the knowledge about the operation, the environment, the cultural constraints of the enemies as well as the allies, the actual rules of engagement, etc. on the one hand side,
- as well as the different work flows of different parties, e.g., the Red Cross, the Croatian Telecom, etc.

The harmonization of all these is very important in OOTW. The military leader is more a director of a campaign he is responsible for, but he isn't able to influence all players directly.

Last but not least the dissemination of information has to be managed as well. It may be crucial to disseminate a piece of information immediately when being available as well as it is possible to take care that some pieces of information should leak. Especially within the work with the media – in a broader sense when doing information operations – the dissemination becomes the most important factor. But also in non information operations it has to be taken into account that the flow of information is not as well hierarchically organized as in „Article V" operations. The national commander may not be the commander having the operational or tactical control, additional sources may have to be informed, etc.

Thus, the following functions are new to the functionality domains when looking at OOTW:

- **Other Systems Presentation**: All information being contained in the systems of the

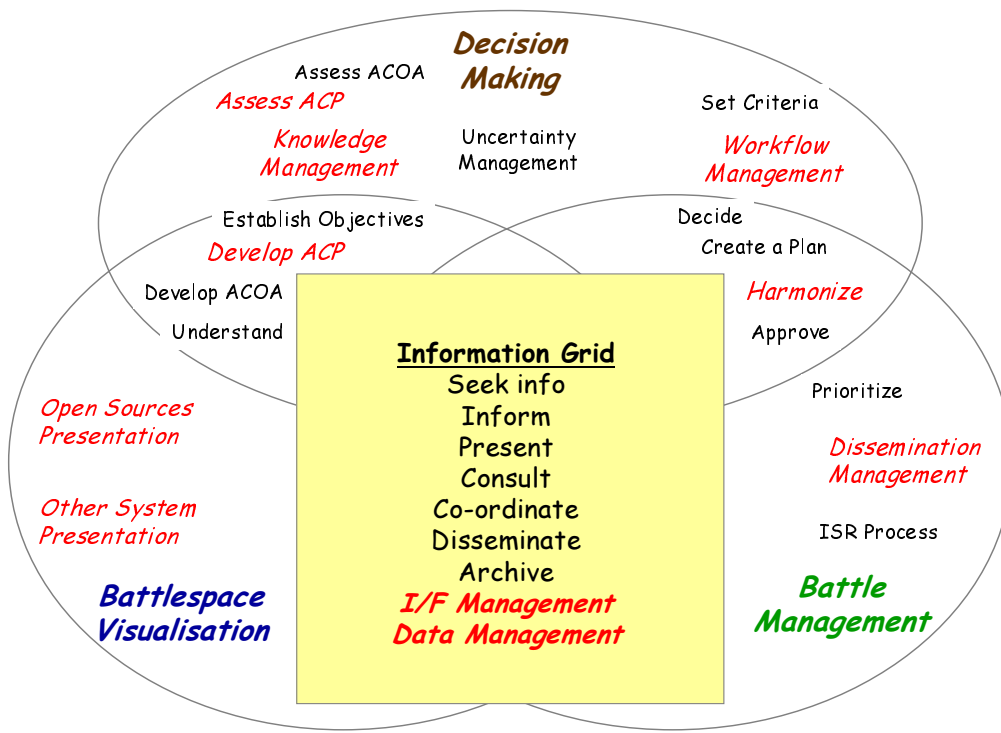

*Fig. 2*: *Function Domains in OOTW*

- partners being relevant to the military decision maker must be assessable as well as presentable.

- **Open Sources Presentation**: The information being available in the open sources have to be assessable and presentable also. Especially the internet and the media have to be mentioned explicitly.

- **Develop Alternative Contingency Plans**: Additionally to ACOA several ACP have to be developed. Especially for ACP their template character have to be taken into account in order to facilitate their access later.

- **Access Alternative Contingency Plans**: Within the given situation, adequate behavior templates in form of preplanned ACP have to be found and presented to the military decision maker in the actual context. In addition to the already introduced ACOA assess, an additional step is necessary before the assessment itself is possible. As the ACP is a template and not a ready to go plan, before assessment becomes possible the respective parameter has to be initialized first. These procedures have to be supported by the underlying C4I system.

- **Knowledge Management**: The knowledge about the situation, the partners, the rules of engagement, potential opponents, cultural constraints on all sides, etc. have to be managed in order to be presented when needed.

- **Workflow Management**: As the workflows of non-military institutions and agency are in general not well known by the decision maker, they have to managed and respective delays to be expected, constraints, hierarchies, etc. have to be made obvious to the commander in a timeliness manner.

- **Harmonize**: The process of harmonizing the decision and respective actions is much more demanding in the non-military environment typical for some OOTW. Thus, additional functionality is needed.

- **Dissemination Management**: Functions supporting the commander and his staff in taking care that the right information reaches the right person or institution in the right time, including consistency management concerning about who have been informed about what and when, is essential to information operations in OOTW and crucial to the overall success of the operation.

Within the information grid, the new function groups for interface management as well as data management have to be established in order to get physically, syntactically, and semantically consistent connected to the various systems of potential partners. This will also be the topic of another section later in this article.

**Using Simulations to Support the Functionality of Future C4I Systems**

It goes without saying that it is not easy to support all these functions by simulation systems. However, especially when looking at the most recent findings of respective research and development studies, there may be some techniques to be useful in the near future. At least the problem of integrating respective applications can be seen as solved in principle (see next section). In order to cope with these issues, we divide potential support by simulation systems into two subsection: educational issues and support to operations.

In general, simulation systems can help to orchestrate the agile management needed in modern warfare by taking into account not only the entities and its relations, but also their dynamical aspects, their behavior, their capabilities as well as their constraints and limitations. Although being just one out of a great variety of possible other means of Operations Research, simulation systems play a special role due to their wide acceptance in different application domains in the armed forces.

*Education and Training: The Need for Computer Assisted Exercises*

It is obvious, that all of these topics having been dealt with in the last section have to be covered by exercises, and therefore should be within the focus of respective CAX systems sooner or later. It should be mentioned, e.g., that actual scenarios for OOTW at, e.g., the Warrior Preparation Center (WPC) at Ramstein (Einsiedlerhof), Germany, are only script driven exercises as we are still lacking respective CAX tools to support efficient training. But even if no special CAX environments are yet available, the training of making use of the respective functions is a must.

As Colonel (ret.) Kenneth Allard – a veteran from the Bosnian operation IFOR – concludes in his contribution to [Wentz, 1998]: *"But the Bosnia experience should also remind us that our worship of technology in warfare must be tempered by a stronger sense of the human factor. Information technology is uniquely affected by people, their training their procedures, and the time they take to perform them. But the combination of these factors in combat and operational settings is constantly and curiously underestimated."*

As already pointed out in the presentation of the findings of respective German studies [Tolk & Kunde, 2000; Tolk & Schiefenbusch, 1999], decision support systems – as well as every supporting information technique - and their use must be integrated as early as possible in training and daily use. What has to be ensured is that the user of decision support systems knows and understands the assumptions and limits of the system and the OR methods to be applied. Decision support systems training means therefore also training in OR methods.

In the eyes of the author respective training of using the supporting functionality therefore is the most crucial element in making this support a success. The need for respective CAX support for OOTW already reached a threshold. Active R&D delivering prototypes are needed as soon as possible.

*Support to Operations: Decision Support Systems*

For some functions it may be possible to use simulation systems as applied means of OR to support the warfighter directly. This approach is followed especially by the US, e.g., with the requirement for integrated tools for ACOA analyses that are part of the future concept for the Army Battle Management System of the future.

Respective simulation systems can be a help in order to make the overall command and control process more agile.

- Simulation systems can help the military decision maker to build up and understand a perceived situation. The main advantage of a simulation is the dynamic of its nature. A simulation system doesn't just display a static picture, but it is possible to include historical data having let to the situation, doing "what-if" analysis, and showing trends or contradictions within the perception. Inconsistencies can be visualized easily too.

- The development of alternative courses of action is, what many analyses tools have been designed for. It just remains the question whether the military users in the field will be able to use this tools, or whether industry support – and it is possible to think of on site support also – is necessary.

- Stochastic combat models can assist in assessing the sensitivity of single parameters as well as the stability of solutions via parametric variation and many replications. How this can be done have already been proven during the NATO study "Stable Defense" by the University of the Federal Armed Forces of Germany, where the simulation model KOSMOS was used to simulate over 60,000 battles to generate quasi-empirical data being needed by the analysts.

- Assessing the ACOA as well as ACP including to approve the findings is very similar to looking for inconsistencies in earlier phases. Again, the dynamical structure of simulations can be of great help.

- Workflow simulation maybe one of the first applications to be integrated, as the techniques already are well known and can help in understanding the procedures on all participating side. It should be pointed out, however, that this is not a typical combat simulation but a new family of simulation to cope with by the analyst as well as the trainers.

- Simulation systems can be used to harmonize the efforts of different allies by showing the partners the effects of a given decision. Simulation systems and their dynamic can be used to visualize the consequences of some action to non-military agencies and institutions, e.g., the necessity of traffic control, harmonized use of bridges and streets, the area of potential collateral damage in case of an engagement, etc.

Despite all these potential application domains it should be stated clearly that the fidelity of respective simulation are tremendous high when being planned to be used as support to operations functionality deliverers. Therefore, it is likely to assume that additional R&D effort is also necessary on this field before first real operational application will emerge. Another aspect is stressed by Seth Bonder in [Bonder, 2000]: Not models produce new insights, but the analysts with 10-15 years of experience in the respective domain. Again, this illuminates the neccessity for extensive OR training for the military users.

**The Role of Standardization**

After having shown what different sorts of applications offering the needed functionality have to be brought together, and having given some first hints where simulation systems may be helpful, the last section deals with some technical aspects of how to bring all these different applications and functions together. It can be stated, that in the opinion of the author this problem field is already solved by the findings of most recent R&D studies.

As neither standardization nor the definition of a general information integration layer is the main focus of this paper, the interested reader is referred to additional publications, e.g., [Hayes, 2000; Hieb & Sprinkle, 2000; Krusche & Tolk, 1999; Krusche & Tolk, 2000; Tolk, 1999b; Tolk, 2000b].

*Data Management and Data Mediation*

As already pointed out in [Krusche and Tolk, 1999], generally each organization in the domain of defense depends on access to information in order to perform its mission. It must create and maintain certain information that is essential to its assigned tasks. Some of this information is private, of no interest to any other organization.

Most organizations, however, produce information that must be shared with others, e.g., operation plans, location and activity of a given unit, information on the logistics, etc. This information must be made available, in a controlled manner, to any authorized user who needs access to it.

At present, almost every defense information infrastructure exists as a collection of heterogeneous, non-integrated systems. This is also true for C3I systems, and – when trying to bring them together in common joint combined operations – the problem of interconnections even increases. This is due to the fact that each organization builds systems to meet its own information requirements, with little concern for satisfying the requirements of others, or of considering in advance the need for information exchange.

If any information exchange takes place, however, as a rule this information exchange is based on *ad hoc* interfaces. The result is an extremely rigid information infrastructure that costs months and millions to be changed or extended, and, which cannot cope with the increasing demand for widely integrated data sharing between multiple mission-related applications and systems. Actual solutions cannot solve these problems, thus, new ways have to be found in the era of joint and combined operations as well as OOTW.

The definition of standard data elements (SDE) required for information exchange, the coordination and control of their implementation and use within systems have to be the central objectives of an overall data management organization. They may not longer be under the responsibility of system managers who's legal and understandable objective is to optimize their system and, logically, neglecting often the requirements of the superimposed federation of systems.

In general, data management is planning, organizing and managing of data by defining and using rules, methods, tools and respective resources to identify, clarify, define and standardize the meaning of data as of their relations. This results in validated standard data elements and relations, which are going to be represented and distributed as a common shared data model.

The overall objective to be reached by introducing a data management is, to coordinate and to control the numerous system projects technically and organizationally, in order to improve the integrity, quality, security and availability of standard data elements.

After having agreed on a common shared data model and the mapping rules for harmonization defined and distributed by the system independent data management organization, data mediation in the sense of automatic translation of system's data into standardized date elements and vice versa becomes possible.

When using an appropriate toolkit,[2] these results can be used to directly configure a software layer interconnecting the data access layers of different systems with heterogeneous data interpretations. It should be pointed out that the data mediation layer is not an isolated technical solution to gain interoperability between different information systems, but is integrated in an overall data management process.

The kernel idea of this integrated interoperability concept can be summarized as follows:
- The data management agency harmonizes the heterogeneous data models of the legacy systems.
- The results are used to configure the data mediation layer that enables the systems to interchange information based on the common shared data model.

Therefore, management and technique are merged to deliver continuous interoperability of systems and applications.

More information concerning this topic can be found in [Krusche et al., 2000].

*Standardization to benefit from international Variety*

A central point of paper [Tolk, 2000b] is to introduce a positive view of the variety of solutions being developed by NATO and PfP nations due to the different national perspective of how to conduct a common operation as well as different interpretations of the analysts trying to support these efforts.

The main role of standardization must be, to deliver the "glue" between all the resulting components delivering the needed functionality. Figure 3 demonstrates the positive role of standardization in this context: Enabling the access of different and variant interpretations of a common multinational effort.

---

[2] *Respective Toolkits have been developed and applied by the IABG Mediation Think Tank in German harmonization projects.*

When bringing different nations (or institutions, agencies, etc.) together to commonly support an operation (or exercise, training program, etc.), there will always be different interpretations of the task, resulting in a first variety of mission interpretations. Analyzing the possibility of support to these interpretations by means of operations research (and using simulation systems is an possibility), the next level of variety is gained.

The result of these process is a wide variety of components supporting the most (and hopefully all) different aspects of possible interpretations of a common operation. This, however, can be seen as a positive synergistic effect, as different views are early supported and therefore the chance to have a support function available if needed increases. The role of standardization in this context is to create the framework being able to comprise and integrate all resulting components.

## Summary

It has been shown, that for NATO operations a wide set of functions and function groups is needed which has to be supported by the underlying C4ISR systems. These functions cover all three domains of C4I application domains: battle visualization, decision making, and battle management. Simulation systems can add support. On the one hand side, they have to support training and education by starting to improve or rebuild actual CAX systems in such a way, that the training of the use all function groups is possible. On the other hand side, simulation systems can be used for support to operations also.

It is assumed, that the use of simulation systems for support to operations implemented as integrated components of the underlying C4I system allows a more agile synchronization in the dynamic environment by involving a dynamic component that can facilitate the orchestration of the many relationships among the dimensions of time, space, arenas, alternatives, etc.

In many cases, the nations will interpret the common mission differently, focussing on different aspects. This variety is increased by different analysts' views or technical concepts. In the eye of the author, these different approaches can be very fertilizing, as different nations focus on different aspects in their doctrines, which results – as a whole – in a broader coverage of functionality really needed for the overall operation. The different doctrines – being harmonized – can support or even complete each other. Therefore, standardization in form of the proposed integrated interoperability approach – aligning management procedures (data management) and technical solutions (data mediation, configurable system interfaces) – is a key factor to benefit from one of the most important advantages of NATO: the wide variety of ideas and visions aiming at the common goal of the alliance.

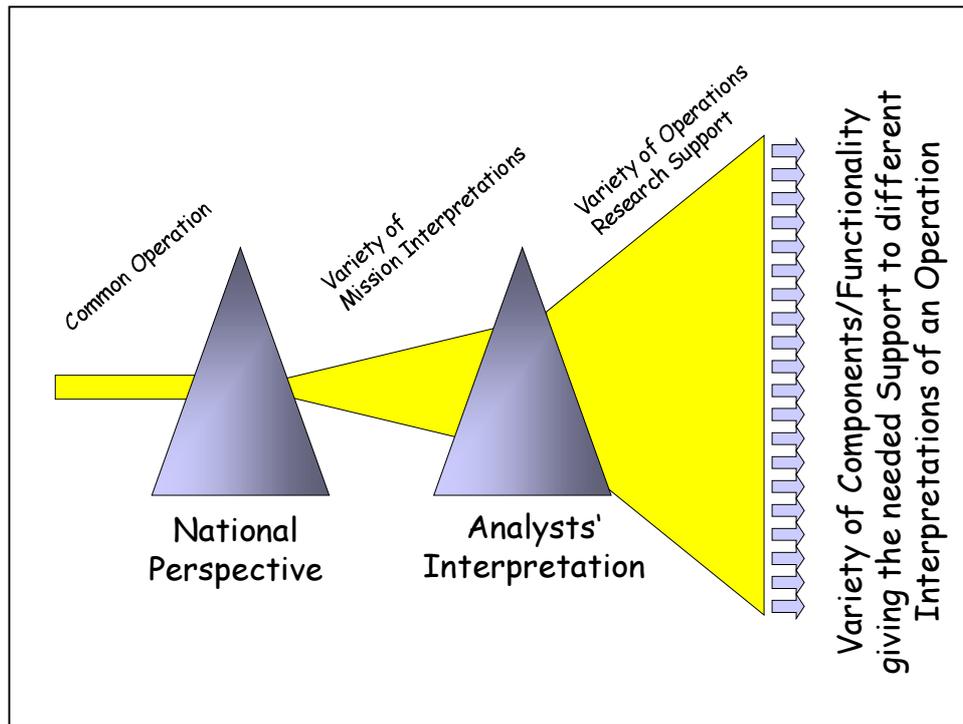

*Fig. 3: Variety of Support resulting from the cascading Variety of different Nations' and Analysts' Interpretations*

**About the Author**

**ANDREAS TOLK** is Project Manager at IABG, Ottobrunn, Germany. He has experiences in architecture and software concepts and design for C4I systems as well as for advanced distributed simulation systems (mainly using HLA). He was a technical expert within the Industry Policy Subgroup (IPSG) helping to formulate the NATO Modeling and Simulation Master Plan. He is Vice Chair of the C4I System Simulation and Integration Forum (SIW-C4I) as well as for the Information Operations and Intelligence, Surveillance, and Reconnaissance Forum (SIW-IO/ISR) of the Simulation Interoperability Standardization Organization (SISO). He is a technical expert within the NATO RSG for C2 Assessment in OOTW. He gives lectures at the University of the Federal Armed Forces of Germany on the Use of Artificial Intelligence within Military Systems and on Data Modeling and Management in Command and Control Systems. He is invited by NATO to give lessons as a Matter-of-Subject expert on interoperability issues within the PfP-Lecture Series.

*[Note: This paper will also be presented at the NATO Regional Conference on Military Communication and Information Systems 2000 (RCMCIS 2000) in Zegrze, Poland, October 2000]*